\documentclass{article} 
\usepackage{iclr2017_conference,times}
\usepackage{hyperref}
\usepackage{url}

\usepackage{amssymb}
\usepackage{amsthm}
\usepackage{stfloats}
\usepackage{amsmath}

\usepackage[]{algorithm2e}
\usepackage{listings}
\usepackage{url}
\usepackage{proof}

\usepackage[utf8]{inputenc} 
\usepackage[T1]{fontenc}    
\usepackage{hyperref}       
\usepackage{url}            
\usepackage{booktabs}       
\usepackage{amsfonts}       
\usepackage{nicefrac}       
\usepackage{microtype}      
\usepackage{graphicx}
 \usepackage{pdfpages}
 \usepackage{subcaption}
 \usepackage{listings}

\usepackage{hyperref}
\usepackage{array}

\title{Fast Chirplet Transform to Enhance CNN Machine Listening - Validation on Animal calls and Speech}

\author{Herv\'e Glotin \\
DYNI, LSIS, Machine Learning \& Bioacoustics team\\
AMU, University of Toulon, ENSAM, CNRS, IUF\\
La Garde, France\\
\texttt{glotin@univ-tln.fr} \\
\And
Julien Ricard\\
DYNI, LSIS, Machine Learning \& Bioacoustics team\\
AMU, University of Toulon, ENSAM, CNRS\\
La Garde, France\\
\texttt{julien.ricard@gmail.com} \\
\And
Randall Balestriero\\
Department of Electrical and Computer Engineering\\
Rice University\\
Houston, TX 77005, USA \\
\texttt{randallbalestriero@gmail.com} \\
\And
}

\begin{document}

\maketitle

\begin{abstract}
The scattering framework offers an optimal hierarchical convolutional decomposition according to its kernels. Convolutional Neural Net (CNN) can be seen as an optimal kernel decomposition, nevertheless it requires large amount of training data to learn its kernels. We propose a trade-off between these two approaches: a Chirplet kernel as an efficient Q constant bioacoustic representation to pretrain CNN. First we motivate Chirplet bioinspired auditory representation. Second we give the first algorithm (and code) of a Fast Chirplet Transform (FCT). Third, we demonstrate the computation efficiency of FCT on large environmental data base: months of Orca recordings, and 1000 Birds species from the LifeClef challenge. Fourth, we validate FCT on the vowels subset of the Speech TIMIT dataset. The results show that FCT accelerates CNN when it pretrains low level layers: it reduces training duration by -28\% for birds classification, and by -26\% for vowels classification. Scores are also enhanced by FCT pretraining, with a relative gain of +7.8\% of Mean Average Precision on birds, and +2.3\% of vowel accuracy against raw audio CNN. We conclude on perspectives on tonotopic FCT deep machine listening, and inter-species bioacoustic transfer learning to generalise the representation of animal communication systems.
\end{abstract}

\section{Introduction}

Representation of bioacoustic sequences started with 'Human' speech in the 70'. Speech automatic processing yields to the efficient Mel Filter Cepstral Coefficients (MFCC) representation. Today new bioacoustic representation paradigms arise from environmental monitoring and species classification at weak Signal to Noise Ratio (SNR) and with small amount of data per species.

Several neurobiological evidences suggest that auditory cortex is tuned to complex time varying acoustic features, and consists of several fields that decompose sounds in parallel \citep{kowalski1996analysis,mercado2000modeling}. Therefore it is more than reasonable to investigate the Chirplet time-frequency representation from acoustic and neurophysiological points of view. 

Chirps, or transient amplitude and frequency modulated waveforms, are ubiquitous in nature systems (\cite{flandrin2001}), ranging from bird songs and music, to animal vocalization (frogs, whales) and Speech. 
Moreover the sinusoidal models are a typical attempt
to represent audio signals as a superposition of chirp-like components. Chirp signals are also commonly observed in biosonar systems. 


The Chirplet transform subsumes both Fourier
analysis and wavelet analysis,
providing a broad framework for mapping one-dimensional sound waveforms into a
n-dimensional auditory parameter space. It offers the processing described in different auditory fields, i.e.
cortical regions with systematically related response sensitivities.
Moreover, Chirplet spaces are highly over-complete because there is an infinite number of ways to segment a time-frequency plane, the dictionary is redundant: this corresponds well with the overlapping, parallel signal processing pathways of auditory cortex.

Then we suggest that low level CNN layers shall be pretrained by Chirplet kernels.
Thus, we define and code a Fast Chirplet Transform (FCT). We conduct validation on real recordings of whale and birds, and on Speech (vowels subset of TIMIT). 
We demonstrate that CNN classification benefits from low level layers FCT pretraining. We conclude on the perspectives of tonotopic FCT machine listening and inter-species transfer learning.

\section{Formal definition of Chirplet}
A chirplet can be seen as a complex sinus with increasing or decreasing frequency over time modulated by a Gaussian window to have a localized support in the time and Fourier domain. It is a broad class of filters which includes wavelets and Fourier basis as special cases. As a result, and as presented in \citep{mann1991chirplet,mann1992adaptive}, the Chirplet transform is a generalization of many known time-frequency representations. We first present briefly the wavelet transform framework to extend it to Chirplets.
Given an input signal $x$ one can compute a wavelet transform \citep{mallat1999wavelet} through the application of multiple wavelets $\psi_{\lambda}$. A wavelet is an atom with localized support in time and frequency domain which integrates to $0$. The analytical support of the wavelets is not compact but they are very well localized. It can be considered compact in the applied case where roundoff error lead to 0 quickly after moving around the center frequency.
 The whole filter bank is derived from a mother wavelet $\psi_0$ and a set of dilation coefficients following a geometric progression defined as $\Lambda =\{ 2^{1+j/Q},j=0,...,JQ-1\}$ with $J$ being the number of octave to decompose and $Q$ the number of wavelets per octave. As a result, one can create the filter-bank as the collection $\{\psi_0(\frac{t}{\lambda}):= \psi_{\lambda}, \lambda \in \Lambda \}$.
After application of the filter-bank, one ends up with a time-scale representation, or scalogram, $Ux(\lambda,t):= |(x \star \psi_\lambda)(t)|$ where the complex modulus was applied in order to remove the phase information and contract the space.
It is clear that a wavelet filter-bank is completely characterized by its mother wavelet and the set of scale parameters. Generalizing this framework for Chirplets will be straightforward by now allowing a nonconstant frequency for each filter.
As for wavelets, filters are generated from a Gaussian window determining the time support however the complex sinus has nonconstant frequency over time with center-frequency $f_c$. Since the scope of the parameters leads infinitely many different possible filters, we have to restrain ourselves, and thus create only a fixed Chirplet filter-bank allowing fast computations. The parameters defining these filters include the time position $t_c$, the frequency center $f_c$, the duration $\Delta_t$ and the chirp rate $c$:
\begin{equation}
    g_{t_c,f_c,\log(\Delta t),c}(t)=\frac{1}{\sqrt{\sqrt{\pi}\Delta t }}e^{-\frac{1}{2}\frac{(t-t_c)^2}{\Delta_t^2}}e^{j2\pi(c(t-t_c)^2+f_c(t-t_c))}.
\end{equation}

\section{Proposition of a Fast Chirplet Transform (FCT)}
The parameter space is basically of infinite dimension. Similarly to continuous wavelet transform however, it is possible to use some a priori knowledge in order to create a finite bank-filter. For example, wavelets are generated by knowing the number of wavelets per octave and the number of octave to decompose. As a result, we used the same motivation in order to reduce the number of possible Chirplets required. 
The goal here is not to compute an invertible transform, but rather provide a redundant transformation highlighting transient structures which are not the same tasks as discussed in \citep{coifman1992wavelet,meyer1993wavelets,coifman1994signal}. As a result, we keep the same overall framework as for wavelets with the $Q$ and $J$ parameters. For example parameters for bird songs in this paper are $J=6$ and $Q=16$ with a sampling rate (SR) of $44100$Hz, and $J=4$ and $Q=16$ on speech and Orca with SR=16 kHz).
Finally, since we are interested in frequency modulations, we compute the ascendant and descendant chirp filters as one being the symetrized version of the other.
As a result, we use a more straightforward analytical formula defined with a starting frequency $F_0$, an ending frequency $F_1$, and the usual wavelet like parameters $\sigma$ being the bandwidth. Finally the hyperparameter $p$ defining the polynomial order of the chirp is constant for the whole bank-filter generation. For example, the case $p=1$ leads to a linear chirp, $p=2$ to a quadratic chirp. The starting and ending frequencies are chosen to approximately cover one octave and are directly computed from the $\lambda$ parameters which define the scales.
Finally, following the scattering network inspiration from \citep{bruna2013invariant}, in order to remove unstable noisy pattern, we apply a low-pass filter (a Gaussian blurring) and thus we increase the SNR of the representation.

\begin{align}
    &\Lambda=\{ 2.0^{1+i/Q},i=0,...,J\times Q-1\},\\
    &F_0=\frac{Fs}{ 2 \lambda},\lambda \in \Lambda,\\
    &F_1=\frac{Fs}{ \lambda},\lambda \in \Lambda,\\
    &\sigma  = 2 \frac{d}{\lambda},\lambda \in \Lambda.
\end{align}

\section{Low complexity FCT Algorithm and implementation}

We give here our code of Fast Chirplet Transform (FCT), taking advantage of the a priori knowledge for the filter-bank creation and the fast convolution algorithm
\footnote{We provide our implementation in Annexe and: \url{https://github.com/DYNI-TOULON/fastchirplet.git}}.
Therefore, we first create the Chirplet with the ascendant and descendant versions in once (see Annexe Algo 1).

Then we generate the whole filter-bank (see Algo 2 in annexe) with the defined $\lambda$ and hyper-parameters.

Finally, we use the scattering framework  \citep{bruna2013invariant,anden2014deep}: we apply a local low-pass filter to the obtained representation. In fact, the scattering coefficients $Sx$ result from a time-averaging on the time-frequency representation $Ux$ bringing local and up to global time-invariance. This time-averaging is computed through the application of the $\phi$ filter, usually a Gabor atom with specified standard deviation and such that
\begin{equation}
    \int \phi(t)dt=1.
\end{equation}
As a result, one computes these coefficients as: $Sx(\lambda,t)=\left( |x \star \psi_\lambda|\star \phi \right)(t)$, where $\psi_\lambda$ is a Chirplet with $\lambda$ parameters and $\phi$. Similarly, we perform local time-averaging on the Chirplet representation in the same manner.

\begin{figure}[t]
    \centering
    \includegraphics[width=0.8\linewidth,angle=-90]{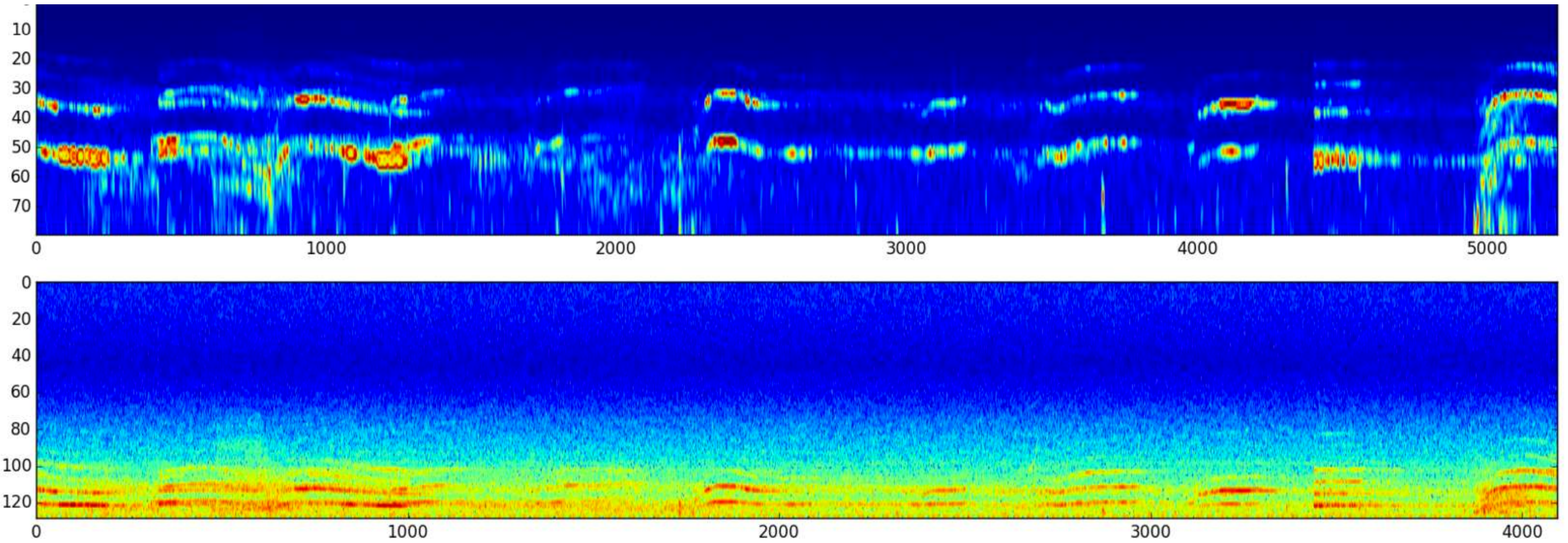}
    \includegraphics[width=0.9\linewidth]{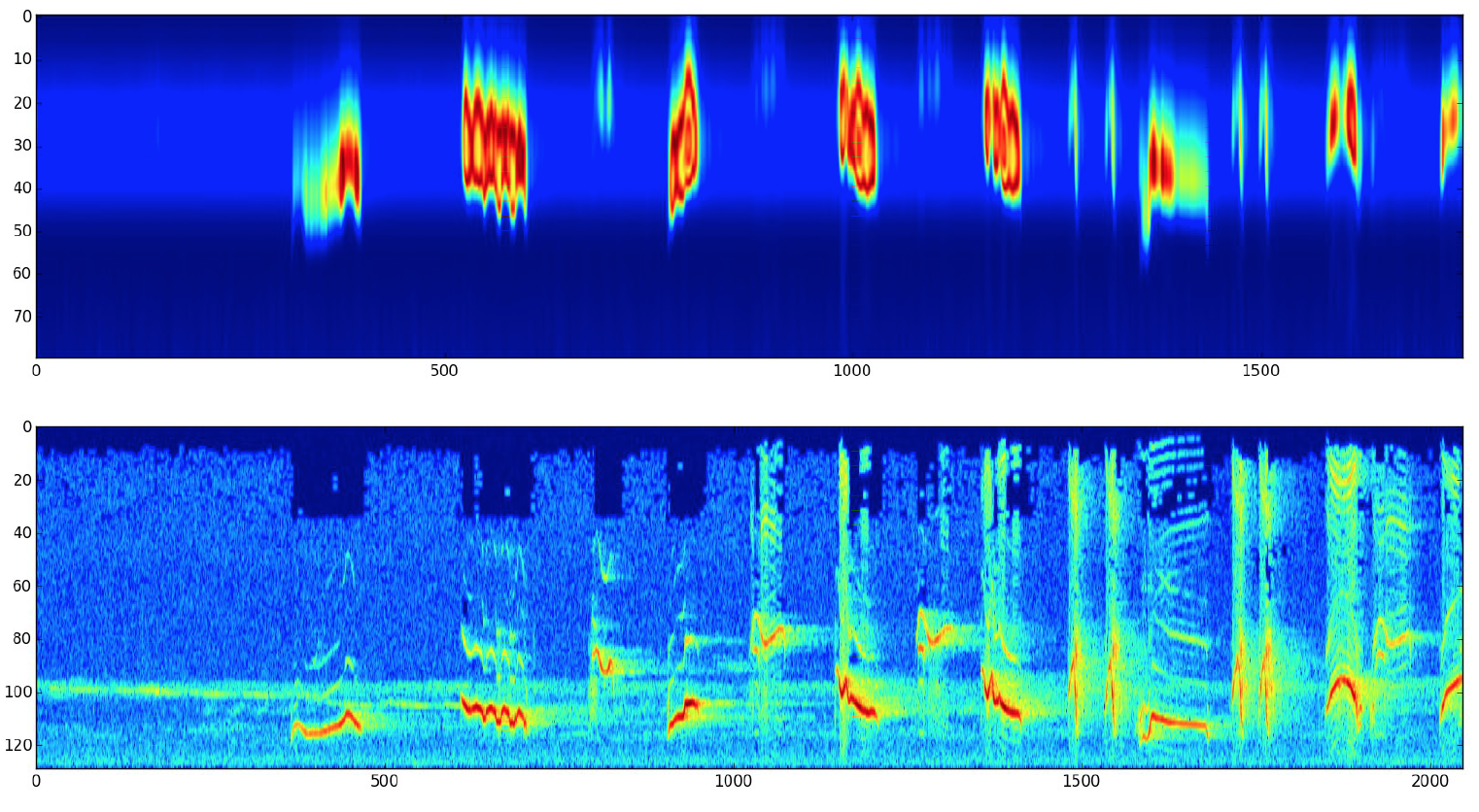}
    \caption{Top: Chirplet of Orca call with p=3, j=4, q=16, t=0.001, s=0.01, with usual FFT spectrogram below, Sampling Rate (SR) 22 kHz, 16 bits. Waves and Chirplets of Orca are: \url{http://sabiod.univ-tln.fr/orcalab}. Bottom: same on bird calls from Amazonia (BIRD10 data set). SR 16 kHz, 16 bits.}
    \label{fig:my_label}
\end{figure}

\begin{figure}[t!]
\centering
\begin{subfigure}[b]{0.45\textwidth}
    \includegraphics[width=\textwidth]{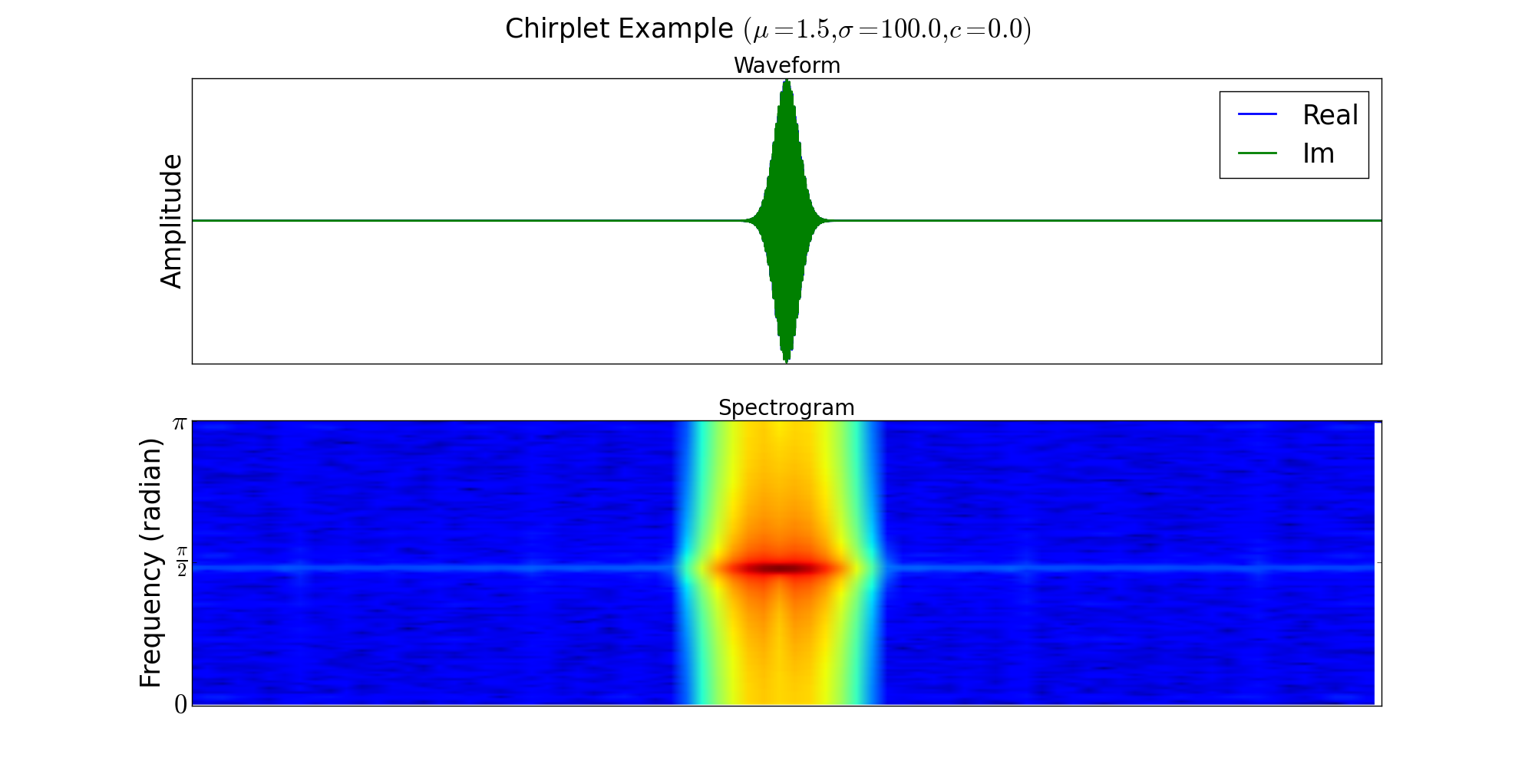}
\end{subfigure}
~
\begin{subfigure}[b]{0.45\textwidth}
    \includegraphics[width=\textwidth]{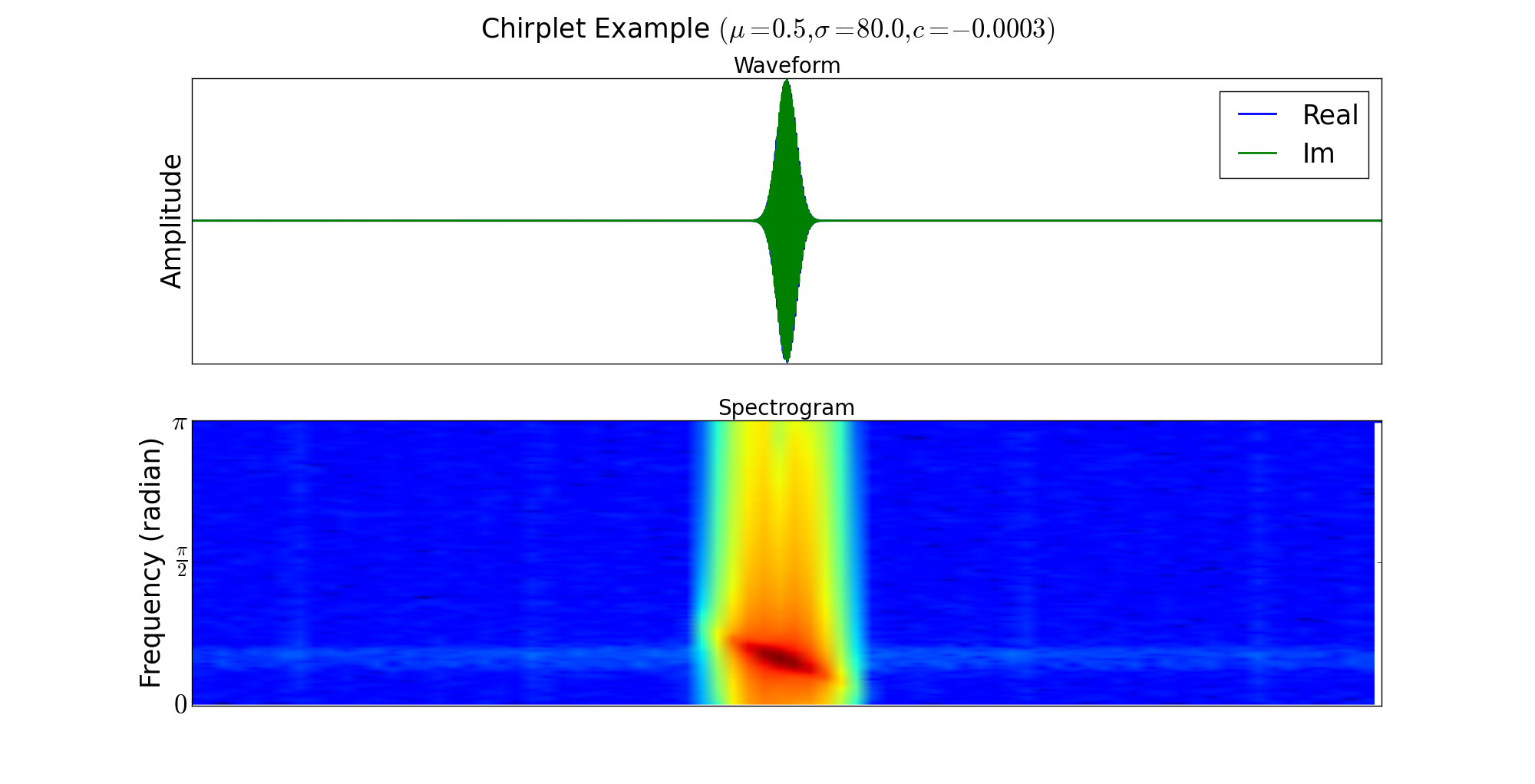}
\end{subfigure}
    ~
\begin{subfigure}[b]{0.45\textwidth}
    \includegraphics[width=\textwidth]{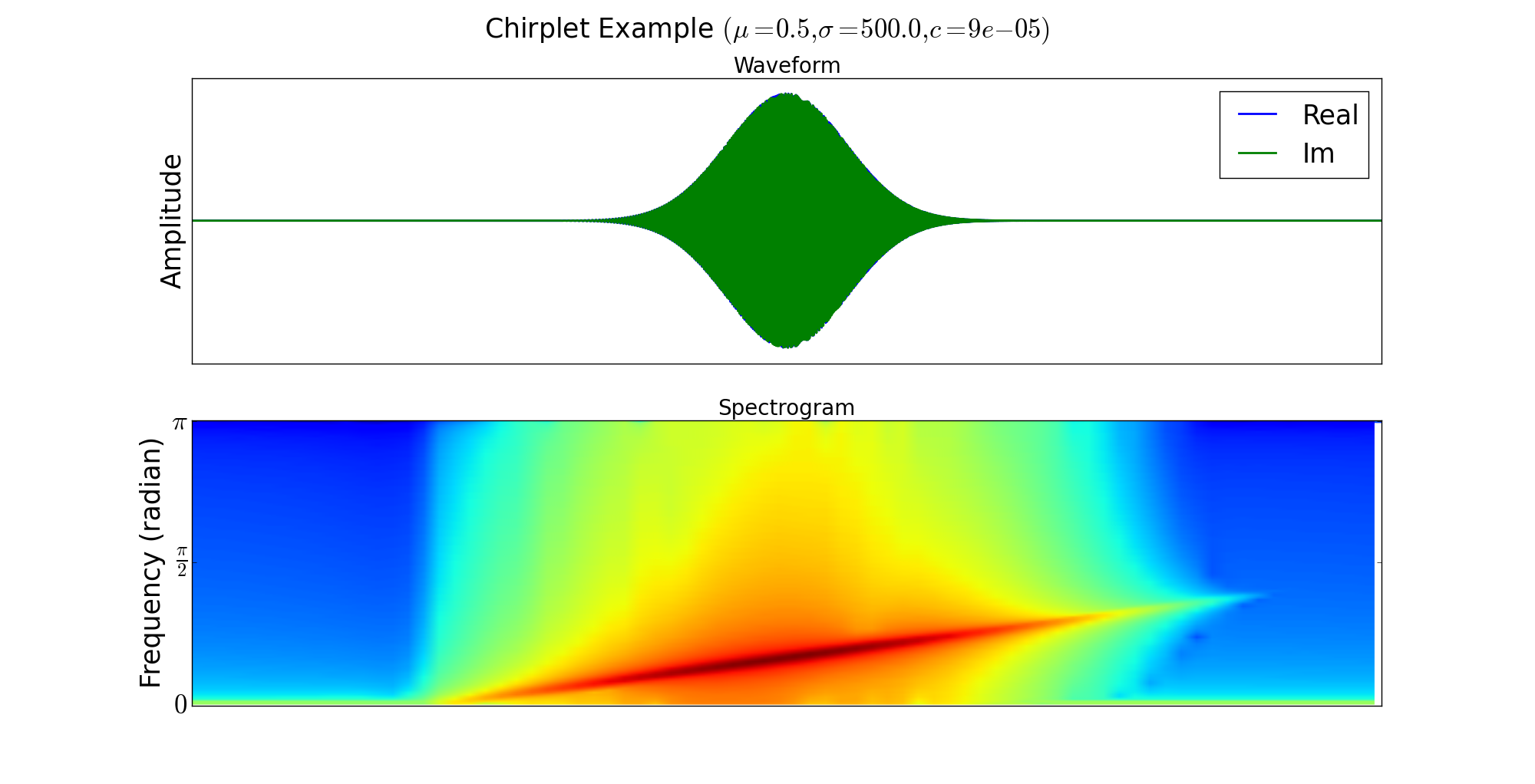}
\end{subfigure}
~
\begin{subfigure}[b]{0.45\textwidth}
    \includegraphics[width=\textwidth]{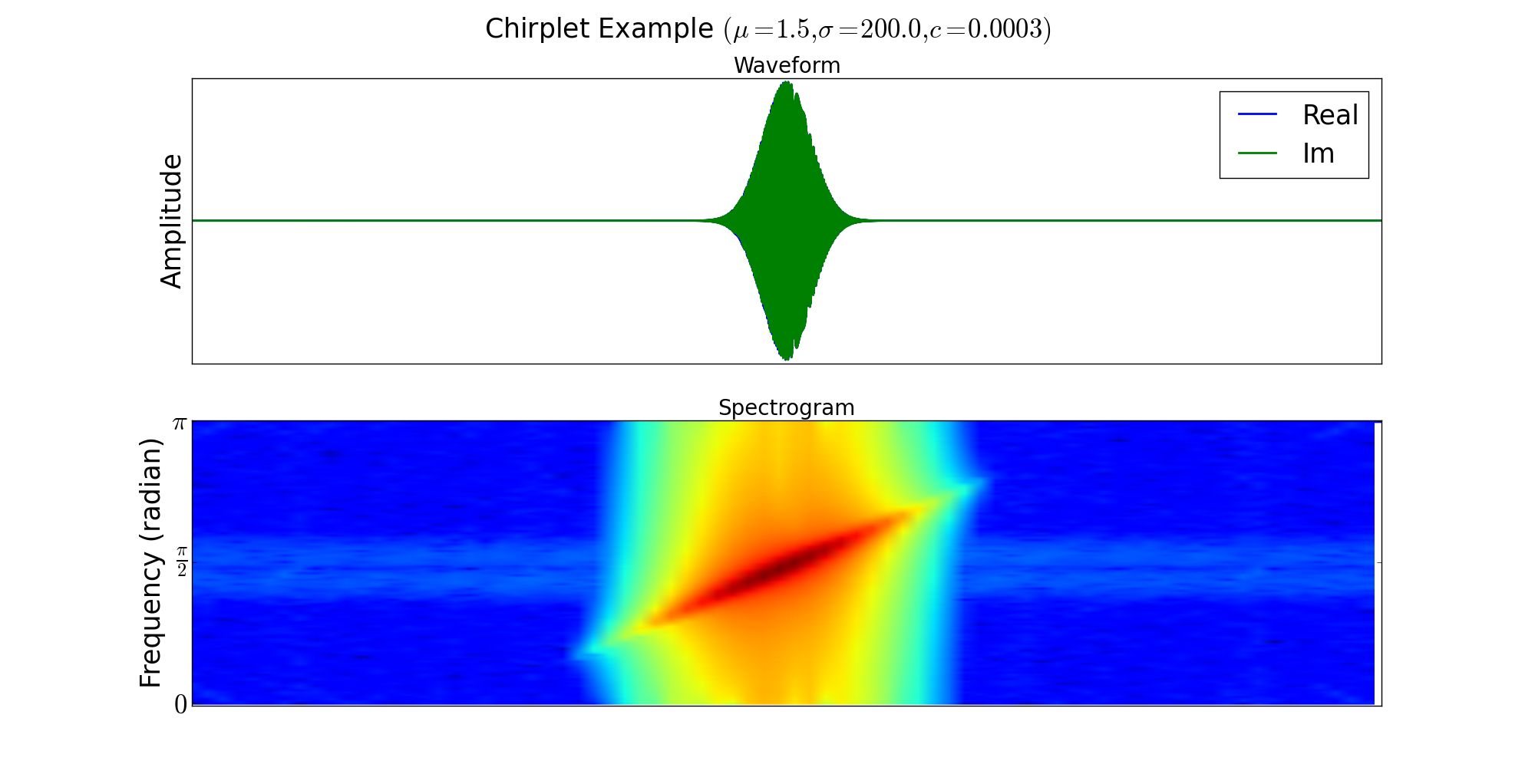}
\end{subfigure}
    ~
\caption{Some FCT displayed in the physical domain and in the time-frequency domain through a spectrogram. The first one reduces to a wavelet since the chirp rate is $0$. One can see the importance of the time duration and the chirp rate and well as the center frequency depending on what one wishes to capture.}.
    \label{fig:chirps}
\end{figure}

\begin{figure}[t!]
\centering
\begin{subfigure}[b]{0.45\textwidth}
    \includegraphics[width=\textwidth]{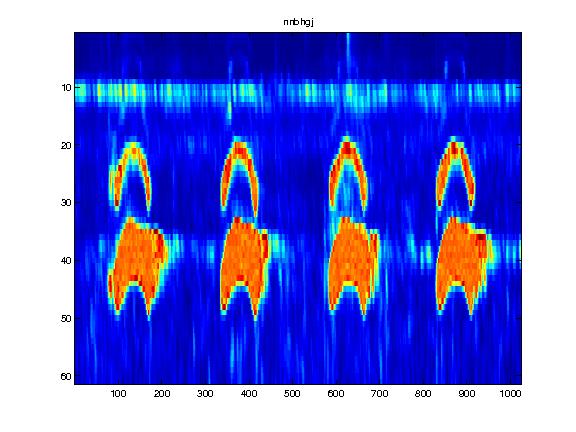}
\end{subfigure}
\begin{subfigure}[b]{0.45\textwidth}
    \includegraphics[width=\textwidth]{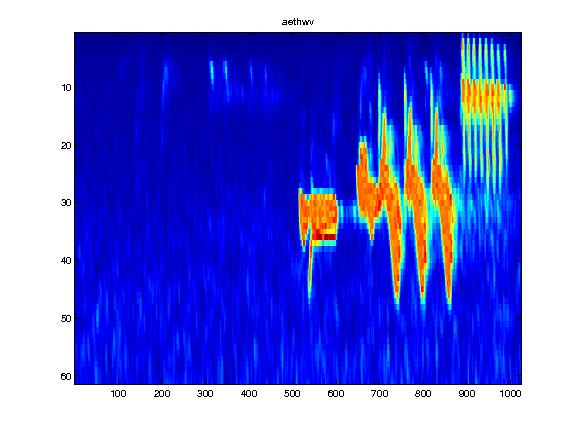}
\end{subfigure}
    ~
\begin{subfigure}[b]{0.45\textwidth}
    \includegraphics[width=\textwidth]{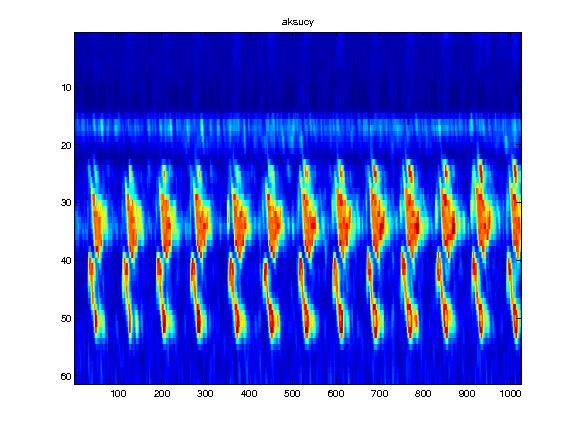}
\end{subfigure}
~
\begin{subfigure}[b]{0.45\textwidth}
    \includegraphics[width=\textwidth]{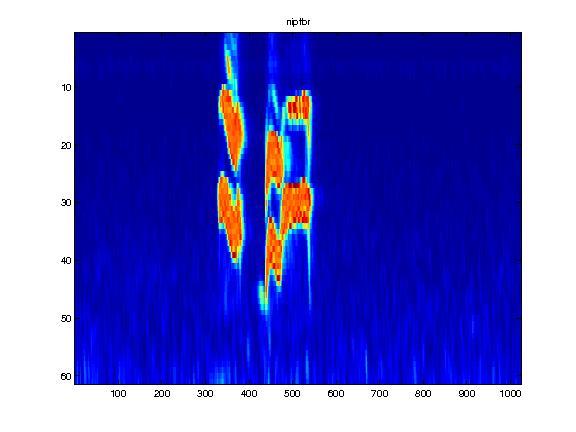}
\end{subfigure}
    ~

\caption{FCT of 4 species of amazonian birds LifeClef 2015 challenge including BIRD10 dataset available online. The call patterns are the high SNR (red) regions. The species international codes are, from top to bottom, right to left: nnbhgj, aethwv, aksucy, nipfbr.}.
    \label{fig:bird}
\end{figure}

We present some possible filters in Fig. \ref{fig:chirps}, and some bird features Fig. \ref{fig:bird}. 

The third step in our FCT consists in the reduction of the convolution task. The asymptotic complexity of the Chirplet transform is $O(N.\log(N))$ with $N$ being the size of the input signal. This is the same asymptotic complexity as for the continuous wavelet transform and the scattering network. However, it is possible to reach lower asymptotic complexity simply by a division of the convolution task. usually the convolutions are carried through application of an element-wise multiplication of the signal and the filter in the frequency domain and then compute the inverse Fourier transform to end up with $x\star \psi_\lambda$.
However, if we denote by $M$ the length of the filter $\psi_\lambda$ it is possible to instead perform multiple times this operation on different overlapping chunks of the signal to then concatenate the results to obtain at the end the same convolution result but now in $O(N.\log(M))$. Finally a last improvement induced by this approach is to allow easy tackling of signals with a length just above a power of $2$ which otherwise would require to be padded in order to obtain a FFT with real $O(N.\log(N))$ complexity through the Danielson-Lanczos lemma \citep{press2007numerical}. Applying this scheme allowed to compute the convolutions between $3$ to $4$ times faster. The variations came from the distance between $N$ and the closest next power of $2$ depending on the desired chunk size.

We validate the efficiency of FCT on real bioacoustic recordings. We processed on 10 medium speed CPUs of 4 years old, 100 hours of recording of LifeClef bird challenge (16 kHz Sampling Rate (SR), 16 bits) in 2 days. Second, we processed in 7 days the equivalent of 1 month of Orca whale recordings from Orcalab.org ONG (22 kHz SR, 16 bits), in Fig. 1,2,3 and at \url{http://sabiod.univ-tln.fr/orcalab} .






\section{Enhancing CNN bioacoustic representation with FCT}
A strategy for CNN fine-tuning can be to retrain a classifier on top of a CNN on a new dataset, or to fine-tune the weights of a pretrained network by continuing the backpropagation. It is possible to fine-tune all the layers of the CNN or to freeze some of the earlier, later or central layers, and to only fine-tune some portion of the network. 
As the features propagate deeper and deeper in the network layers, they become increasingly invariant and discriminative (Seltzer 2013).
Thus usually only the higher level are fine-tuned, the earlier features of a CNN contain more generic features that should be useful to many tasks. As denoted in later layers of the CNN becomes progressively more specific to the details of the classes contained in the original dataset.


In this paper we adapt our parametric Chirplet decomposition to a specific acoustic domain with a specific CNN. We compare a CNN trained on raw audio to one trained on Mel and Chirplet. The best model is the one trained on parametric Chirplet. Second, we show that the CNN can be enhanced by pretraining Chirp in low level layer.

\subsection{CNN Birds classification on FCT, raw audio, versus Mel}

The first demonstration is conducted on complex Bird songs. We use the BIRD10 subset of LifeClef 2016 bird classification challenge. It was used as ENS Ulm data challenge 2016, and contains 3 species in a total of 15 minutes of recordings (SR 44100 Hz, 16 bits), and is available (.wav, Mel and FCT features) at \url{http://sabiod.univ-tln.fr/workspace/BIRD10}.

We train 3 CNNs \citep{lecun1995convolutional} on the Lasagne Theano platform.
The baseline CNN is trained from the raw audio. A second CNN, with similar topology (see annexe) is trained on a simple log of the simple 64 channels Mel scale of FFT spectrum ( http://pydoc.net/Python/librosa/0.2.0/librosa.feature/ ). We overlap by 90\% the time windows.
A third CNN is trained on our FCT. The parameters of both CNN are similar, with 64 frequency bands each (we remove top and bottom band from the Chirplet to set to 64 bands only). Then the 
input layer is 64 x 86, the Conv layer has 20 filters of size 8 x 10.
All activation functions are relu. We maxpool 2 x 2, follow the 20 filters of size 8 x 10, maxpooling, dense layer (200), dropout at 10\%, with a final softmax dense layer with 3 classes and same dropout. Each CNN is trained by cross-entropy, L2 reg., with a learning rate set of 0.001.

The Fig. 4 gives the MAP of these two CNNs having similar hyperparameters. The CNN on FCT gives the best MAP with 61.5\% at epoch 280 compared to later epoch (820) for Mel with a similar MAP of 61\%. Audio is slower and weaker (58\% MAP at epoch 1140).

\begin{figure}[h!]
    \centering
    \includegraphics[width=1\linewidth]{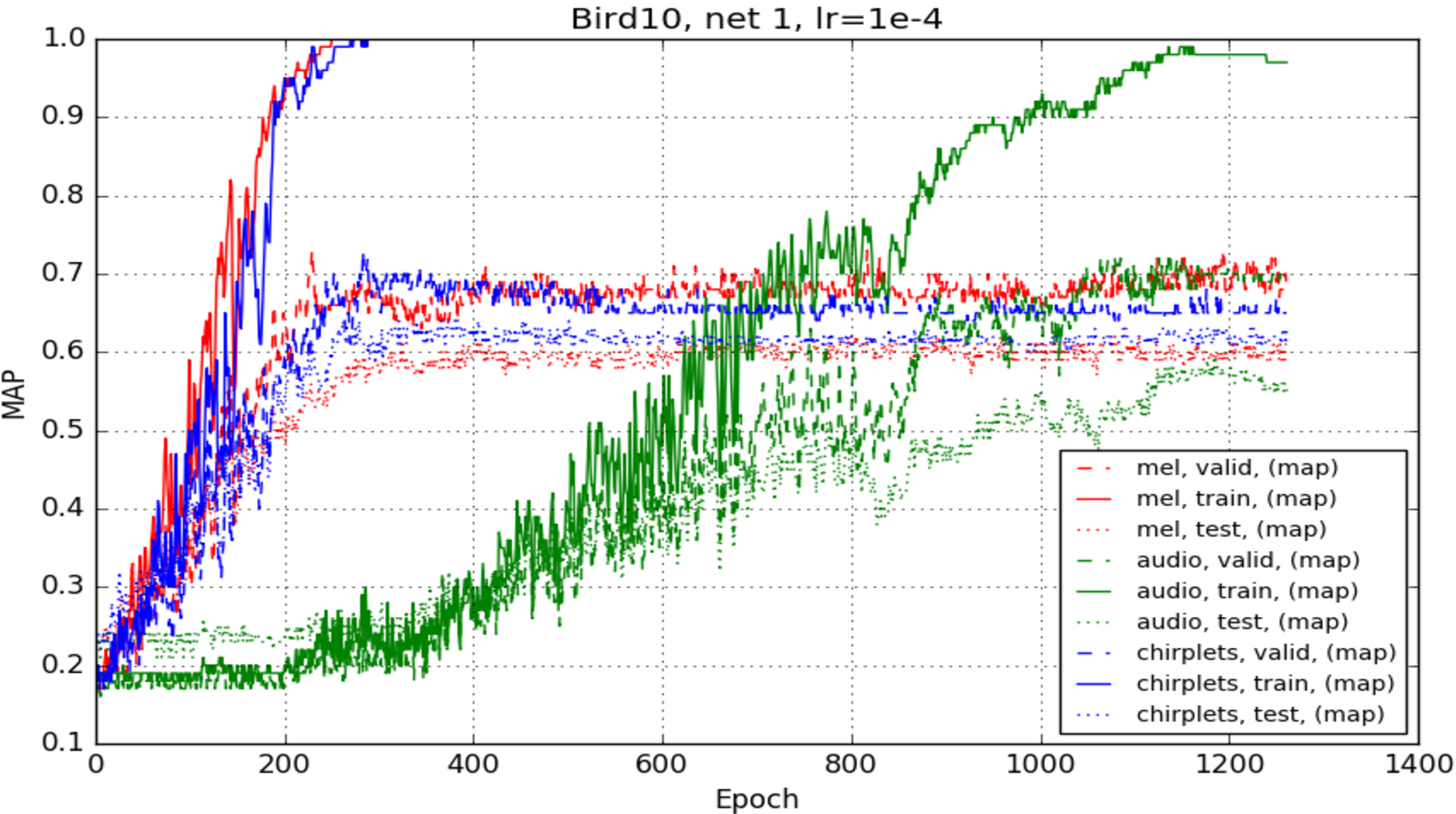}
    \caption{The Mean Average Precision on BIRD10 of the CNNs on Mel, raw audio, or FCT. The training conditions are the same on the three CNNs, and they have similar size and topology (see Annexe). The CNN trained on FCT is slightly better than on Mel or raw audio, and is learning faster.}
    \label{fig:my_label3}
\end{figure}

\subsection{Enhancing Birds classification stacking pretrained Chirpnet CNN}

In order to test the efficiency of the FCT, we pretrain a CNN to encode audio to Chirplets (a.k.a. the Audio2Chirp CNN) and a CNN to convert parametric Chirplet to classes (a.k.a. the Chirp2Class CNN).
The topology of these CNNs (Tab. 2, 3) is set for reasonable time of training. We also speed up the training with shorter time overlap of the time windows (only 30\% instead of 90\% in the previous experimentation). We then decrease the average MAP, however the objective here is to compare the gain in MAP and time of convergence in stacked Chirplet deep representations.

We then simply stack at low level layer the audio2chirp with the chirp2class CNN to build a complete audio2class CNN. We train it from random initialization, or from pretrained CNN. Note that the random seed in all the experimentation of this paper is fixed to allow fair comparisons.
Results are reported in Tab. 1 for each of the stacked CNN, with the epoch giving the best MAP on the dev. set, and the corresponding MAP on the test set. Results demonstrate that the pretraining of low level layers by FCT enhances CNN. More details are given in Annexe.

\begin{center}
\begin{table}[h!]
  \begin{tabular}{| l | c | c |}
    \hline
    \textbf{Model} & BIRD dev & BIRD test  \\   
 & \textbf{nb. Epoch}  & \textbf{MAP \%}   \\ \hline
 
    Baseline: & &\\
    Audio2Class(0) CNN trained from random initialisation (Fig. 6) & 530 (rel.gain) & 51 (rel. gain)  \\ \hline
    
    
    Audio2Class CNN from stacked & &   \\ 
    pretrained Audio2Chirp(*) with Chirp2Class(*) CNNs (Fig. 7) & 390 (-26\%) & 53 (+3.9\%)  \\ \hline

Audio2Class CNN from stacked  & &\\   
pretrained Audio2Chirp(*) with Chirp2Class(*) CNNs   & & \\
without updating Chirp2Class(*) (Fig. 8)  & 380 (-28\%) & 55 (+7.8\%)\\ \hline

\end{tabular}
  \caption {Summary of the CNN enhanced by our FCT representation, on BIRD. For each model, we detail the time of convergence on dev. and corresponding Mean Average Precision on test set.}
  \end{table}
\end{center}

\subsection{Enhancing Vowels classification stacking pretrained Chirpnet CNN}

In this section we run the same demonstration on the subset of speech vowels of all the TIMIT acoustic-phonetic corpus \cite{TIMIT93}: 3,696 training
utterances (sampled at 16kHz) from 462 speakers. The cross-validation set consists of 400 utterances from
50 speakers. The core test set of the 8 vowels subset was used to report the results: 192 utterances from 24 speakers, excluding the validation set. There are 61 hand labeled phonetic symbols but the experiments in this paper run on the time windows of 310ms centered on each of the 8 vowels of TIMIT (= \/iy\/, \/ih\/, \/eh\/, \/ae\/, \/aa\/, \/ah\/, \/uh\/, \/uw\/).

Due to similar bioacoustic voicing dynamics of the two species (near 4 Hz), we simply set the FCT parameters for vowel to the one used for Orca presented above ($p = 3$,$j = 4$,$q = 16$,$t = 0.001$,$s = 0.01$). The time windows are set to 310 ms as recommended in \cite{Collobert2013}.

\begin{figure}[t!]
\centering
\begin{subfigure}[h!]{\textwidth}
    \includegraphics[width=\textwidth]{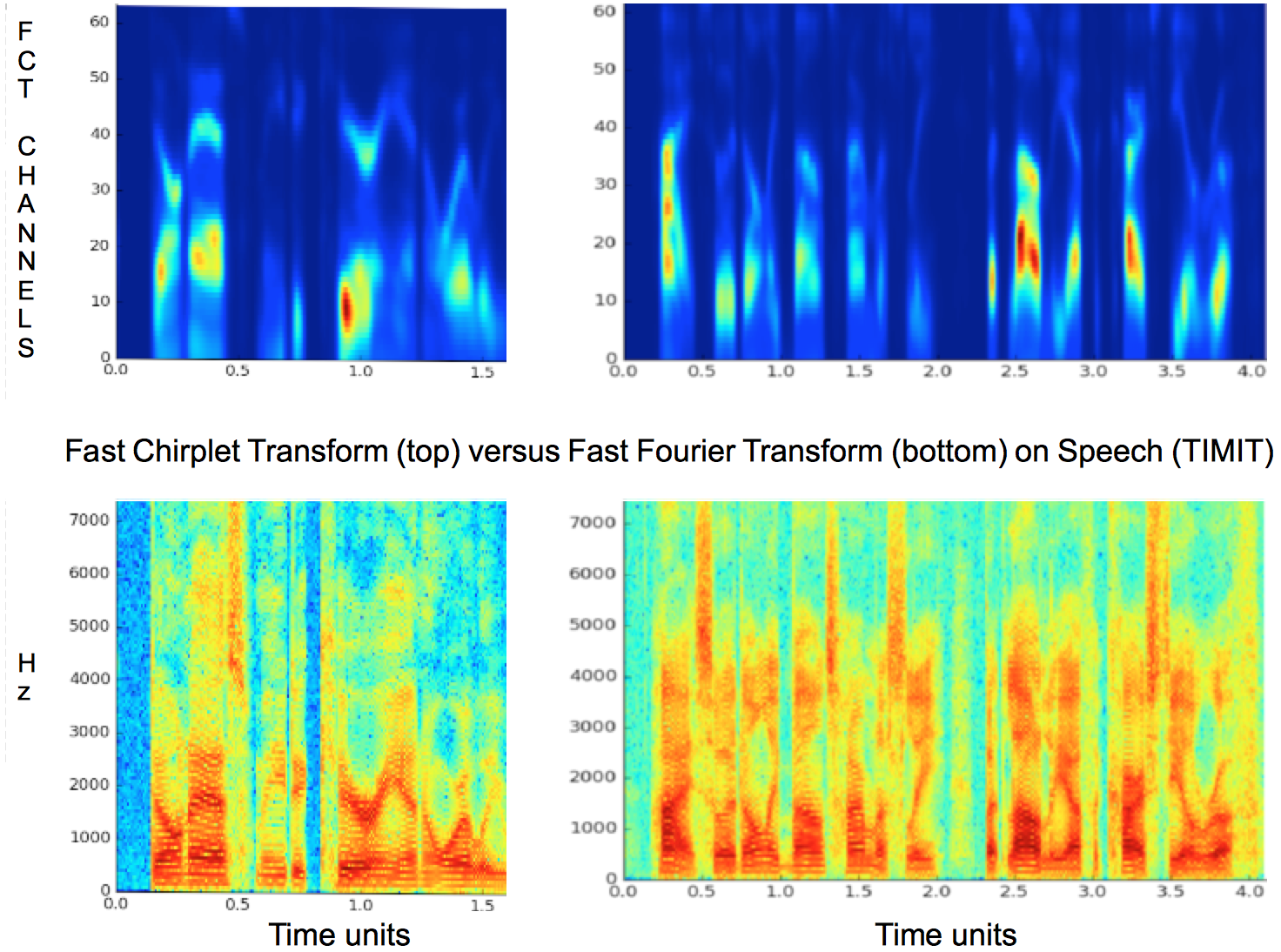}
\end{subfigure}
\caption{FCT (top) versus Fourier spectrogram (bottom) of two utterances of Speech vowel (TIMIT), ($p = 3$,$j = 4$,$q = 16$,$t = 0.001$,$s = 0.01$).}.
    \label{fig:TIMIT}
\end{figure}

The results of the different training stages of the audio2chirp and chirp2class and stacked model are given in Tab.2 and Annexe.
We run due to lack of time the experiment only on vowel classification, which does not really allow comparison with other papers, however this seminal work only aims to study the relative gain between CNN pretrained or not by FCT.

The results demonstrate that FCT pretraining of the audio2class model 
is improved by 2.3\% of relative gain of accuracy while the training time is decreased by 26\%.

\begin{figure}[t!]
    \centering
    \includegraphics[width=0.9\linewidth]{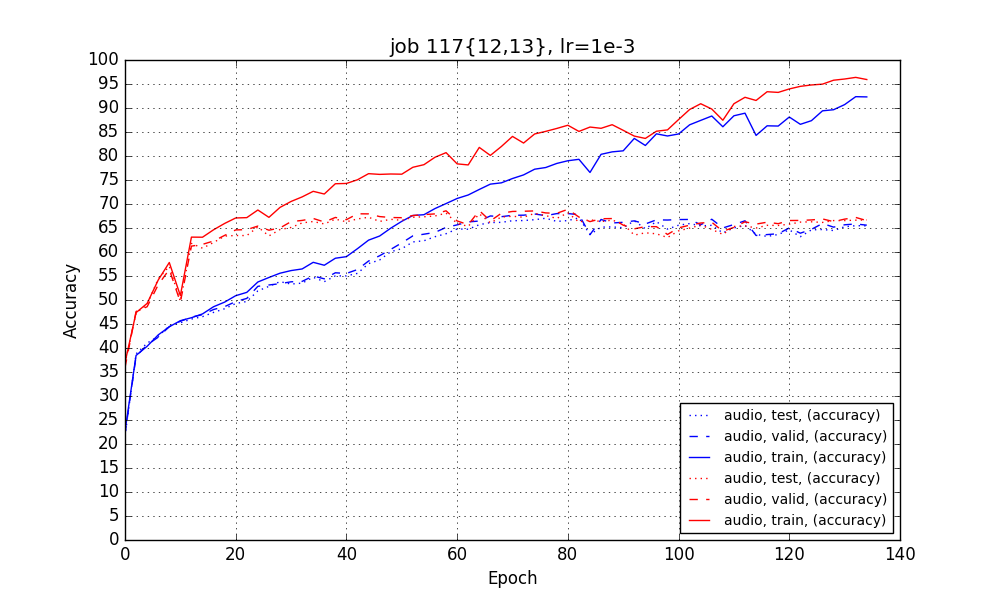}
    \caption{Training stacked CNN. Blue: random initialization of audio2chirp(0) and chirp2class(0). Red: Initialization with optimal audio2chirp(*) and chirp2class(*).}
    \label{fig:my_labe12l3}
\end{figure}

\begin{center}
\begin{table}[h!]
  \begin{tabular}{| l | l | l |}
    \hline
    \textbf{Model} & TIMIT dev & TIMIT test \\
Scores on the 8 vowels of TIMIT & \textbf{nb. Epoch (rel. gain)}  & \textbf{AC \% (rel. gain)}  \\ \hline
    Baseline: Audio2Class CNN from scratch & 79 (ref.) & 66.5 (ref.)\\ \hline
    
    Audio2Class CNN trained from stacked  & &  \\ 
    pretrained Audio2Chirp(*) with Chirp2Class(*) CNNs  & 58 (-26\%) & 68.0 (+2.3\%)  \\ \hline
    
    

\end{tabular}
  \caption {Summary of the CNN enhanced by our FCT representation on vowel (TIMIT): time of convergence and vowel accuracy on TIMIT test set.}
  \end{table}
\end{center}

\section{Discussion and Conclusion}

In this paper we propose for the first time at our knowledge the definition and implementation of a Fast Chirplet Transform (FCT). Due to its low complexity, FCT can be computed as fast as FFT.

Second we show that FCT pretraining accelerates CNN. For Bird10 data set, we have 280 epochs using FCT versus 820 on Mel features, or 1140 on raw audio for same MAP score.
The stacked CNN with the chirpnet in low level layer also decreases training from 530 epochs to 380 epochs, while it increases MAP by 4 points (Tab. 1).
The experiment on Vowels demonstrates a training of 30 epochs on FCT, versus 60 on raw audio (for same 65\% accuracy level), and an increase of 1.5 point of accuracy (Tab. 2).

These gains may be due to the sparsity of the Chirplet, and the denoising step in the FCT.
These experiences bring to light the problem of deep learning for small and biased dataset for which a full learning strategy is sub-optimal due to local optimum convergence. As a result, FCT prior knowledge can be used to mitigate this drawback by reducing the complexity of the deep-net architecture.

Three main perspectives are then opened.
 Future work will consist on sparse Chirpnet inspired from tonotopic net \cite{strom1997tonotopic}, auditory nerve and cortex topology \cite{PironkovDD15}. The acoustic vibrations are transmitted to the base of the cochlea, thus each region of the basilar membrane are excited by different frequencies. The higher frequencies excite areas closer to the cochlea base, whereas lower frequencies are closer to the apex. This implies that neurons connected to a specific zone of the basilar membrane will be simultaneously stimulated inducing tonotopic representation.

A second perspective is to integrate Chirplet computation into the CNN training itself, as a constrained embedded layer, in a framework similar to a Wavelet Neural Network \citep{adeli2006dynamic} but with Chirplet activation functions.

Last, we currently work on transfer learning of Chirpnet from animal to speech (and reverse), in order to generalize a deep Chirpnet representation of the animal communication systems.

\section{Acknowledgements}
We thank colleagues from ENS Paris Data Team with S. Mallat, and P. Flandrin, for fruitful discussions on Scattering and Chirplet. We thank YLC and YB for advises on CNN. We thank V. Tassan for cleaning the code. We used Theano, Lasagne \footnote{https://github.com/Lasagne/Lasagne}, Librosa \footnote{https://github.com/librosa/librosa} and Pysoundfile \footnote{https://github.com/bastibe/PySoundFile}.

\bibliography{iclr2017_conference}

\begin{thebibliography}{18}
\providecommand{\natexlab}[1]{#1}
\providecommand{\url}[1]{\texttt{#1}}
\expandafter\ifx\csname urlstyle\endcsname\relax
  \providecommand{\doi}[1]{doi: #1}\else
  \providecommand{\doi}{doi: \begingroup \urlstyle{rm}\Url}\fi

\bibitem[Adeli \& Jiang(2006)Adeli and Jiang]{adeli2006dynamic}
Hojjat Adeli and Xiaomo Jiang.
\newblock Dynamic fuzzy wavelet neural network model for structural system
  identification.
\newblock \emph{Journal of Structural Engineering}, 132\penalty0 (1):\penalty0
  102--111, 2006.

\bibitem[And{\'e}n \& Mallat(2014)And{\'e}n and Mallat]{anden2014deep}
Joakim And{\'e}n and St{\'e}phane Mallat.
\newblock Deep scattering spectrum.
\newblock \emph{IEEE Transactions on Signal Processing}, 62\penalty0
  (16):\penalty0 4114--4128, 2014.

\bibitem[Bruna \& Mallat(2013)Bruna and Mallat]{bruna2013invariant}
Joan Bruna and St{\'e}phane Mallat.
\newblock Invariant scattering convolution networks.
\newblock \emph{IEEE transactions on pattern analysis and machine
  intelligence}, 35\penalty0 (8):\penalty0 1872--1886, 2013.

\bibitem[Coifman et~al.(1992)Coifman, Meyer, and
  Wickerhauser]{coifman1992wavelet}
Ronald~R Coifman, Yves Meyer, and Victor Wickerhauser.
\newblock Wavelet analysis and signal processing.
\newblock In \emph{In Wavelets and their Applications}. Citeseer, 1992.

\bibitem[Coifman et~al.(1994)Coifman, Meyer, Quake, and
  Wickerhauser]{coifman1994signal}
Ronald~R Coifman, Yves Meyer, Steven Quake, and M~Victor Wickerhauser.
\newblock Signal processing and compression with wavelet packets.
\newblock In \emph{Wavelets and their applications}, pp.\  363--379. Springer,
  1994.

\bibitem[Flandrin(2001)]{flandrin2001}
Patrick Flandrin.
\newblock Time frequency and chirps.
\newblock \emph{Proc. SPIE}, 4391:\penalty0 161--175, 2001.
\newblock \doi{10.1117/12.421196}.
\newblock URL \url{http://dx.doi.org/10.1117/12.421196}.

\bibitem[JS et~al.(1993)JS, Lamel, and al.]{TIMIT93}
Garofolo JS, LF~Lamel, and al.
\newblock Timit acoustic-phonetic continuous speech corpus.
\newblock In \emph{Linguistic data consortium, Philadelphia}, 1993.

\bibitem[Kowalski et~al.(1996)Kowalski, Depireux, and
  Shamma]{kowalski1996analysis}
Nina Kowalski, Didier~A Depireux, and Shihab~A Shamma.
\newblock Analysis of dynamic spectra in ferret primary auditory cortex. ii.
  prediction of unit responses to arbitrary dynamic spectra.
\newblock \emph{Journal of Neurophysiology}, 76\penalty0 (5):\penalty0
  3524--3534, 1996.

\bibitem[LeCun \& Bengio(1995)LeCun and Bengio]{lecun1995convolutional}
Yann LeCun and Yoshua Bengio.
\newblock Convolutional networks for images, speech, and time series.
\newblock \emph{The handbook of brain theory and neural networks},
  3361\penalty0 (10):\penalty0 1995, 1995.

\bibitem[Mallat(1999)]{mallat1999wavelet}
St{\'e}phane Mallat.
\newblock \emph{A wavelet tour of signal processing}.
\newblock Academic press, 1999.

\bibitem[Mann \& Haykin(1991)Mann and Haykin]{mann1991chirplet}
Steve Mann and Simon Haykin.
\newblock The chirplet transform: A generalization of gabor’s logon
  transform.
\newblock In \emph{Vision Interface}, volume~91, pp.\  205--212, 1991.

\bibitem[Mann \& Haykin(1992)Mann and Haykin]{mann1992adaptive}
Steve Mann and Simon Haykin.
\newblock Adaptive chirplet transform: an adaptive generalization of the
  wavelet transform.
\newblock \emph{Optical Engineering}, 31\penalty0 (6):\penalty0 1243--1256,
  1992.

\bibitem[Mercado et~al.(2000)Mercado, Myers, and Gluck]{mercado2000modeling}
Eduardo Mercado, Catherine~E Myers, and Mark~A Gluck.
\newblock Modeling auditory cortical processing as an adaptive chirplet
  transform.
\newblock \emph{Neurocomputing}, 32:\penalty0 913--919, 2000.

\bibitem[Meyer(1993)]{meyer1993wavelets}
Yves Meyer.
\newblock Wavelets-algorithms and applications.
\newblock \emph{Wavelets-Algorithms and applications Society for Industrial and
  Applied Mathematics Translation., 142 p.}, 1, 1993.

\bibitem[Palaz et~al.(2013)Palaz, Collobert, and Magimai{-}Doss]{Collobert2013}
Dimitri Palaz, Ronan Collobert, and Mathew Magimai{-}Doss.
\newblock Estimating phoneme class conditional probabilities from raw speech
  signal using convolutional neural networks.
\newblock \emph{CoRR}, abs/1304.1018, 2013.
\newblock URL \url{http://arxiv.org/abs/1304.1018}.

\bibitem[Pironkov et~al.(2015)Pironkov, Dupont, and Dutoit]{PironkovDD15}
Gueorgui Pironkov, St{\'{e}}phane Dupont, and Thierry Dutoit.
\newblock Investigating sparse deep neural networks for speech recognition.
\newblock In \emph{IEEE ASRU Workshop}, pp.\  124--129, 2015.

\bibitem[Press(2007)]{press2007numerical}
William~H Press.
\newblock \emph{Numerical recipes 3rd edition: The art of scientific
  computing}.
\newblock Cambridge university press, 2007.

\bibitem[Strom(1997)]{strom1997tonotopic}
Nikko Strom.
\newblock A tonotopic artificial neural network architecture for phoneme
  probability estimation.
\newblock In \emph{Automatic Speech Rec. and Understanding IEEE Wkp}, pp.\
  156--163, 1997.

\end{thebibliography}
\bibliographystyle{iclr2017_conference}

\newpage

\appendix  
  
\section{BIRD Dataset}
The experiment is conducted on BIRD10, an online data set \url{http://sabiod.univ-tln.fr/workspace/BIRD10/} which is a subset of the training LIFEClef 2016 challenge on bird classification. BIRD10 contains 454 audio files (22050 Hz SR, 16 bits) from 10 bird classes, split in 0.5s segments. 20\% of the training set was used as the validation set.

Only segments with detected bird activity were kept, assuming a bird sound to have prominent energy and to be mostly harmonic. This bird detection is for a given segment:

\begin{verbatim}
    if (energy_ratio > energy_threshold and
        spectral_flatness_weighted_mean < spectral_flatness_threshold)
        bird_detected = True
    else
        bird_detected = False
\end{verbatim}

where the energy and the spectral flatness are computed on 50\% overlapping frames of 256 samples:
$$er=energy\_ratio=\frac{mean(seg\_energy)}{95thpercentile(file\_energy)}$$
$$swf=spectral\_flatness\_weighted\_mean=\frac{sum(seg\_spectral\_flatness\times seg\_energy)}{sum(seg\_energy)}$$
This naive algorithm performed quite well on a manually labelled dataset of bird vocalizations (precision=0.89, recall=0.57 for er=0.2 and sfw=0.3) after a quick grid search on the two parameters.

\section{BIRDs classification : baseline CNNs}

The first experiment consisted in running similar CNNs to compare the performance of using raw audio and two time-frequency representations as the input: a standard log-amplitude Mel spectrum and the Chirplet representation described in the first part of this paper. In this experiments the segments were overlapping by 90\%. The topologies of the networks are given in Tab. 2. The cost function is the cross-entropy, learning rate = 0.0001 = L2 regularisation coefficient. 
The Mel spectrum is computed from 64 bands between 0 and 11025 Hz (=SR/2).
Both Mel spectrum and Chirplets were normalized by Z-score.

\begin{center}
\begin{table}[h!]
  \begin{tabular}{| c | c |}
    \hline
    \textbf{Input} & \textbf{Topology} \\ \hline
    \textbf{Audio}, shape (1, 11025) &
    \begin{tabular}[x]{l@{}l@{}l@{}l@{}l@{}l@{}}conv\_1: 20 filters of shape (1, 400) (nonlinearity: relu)\\
    pool\_1: (1, 4) max pooling\\
    conv\_2: 20 filters of shape (1, 100) (nonlinearity: relu)\\
    pool\_2: (1, 4) max pooling\\
    dense\_1: 400 units (nonlinearity: relu, 10\% dropout)\\
    dense\_2: 10 units (nonlinearity: softmax, 10\% dropout)    
    \end{tabular} \\ \hline
    \textbf{Log-amplitude Mel spectrum}, shape (64, 80) &
    \begin{tabular}[x]{l@{}l@{}l@{}l@{}l@{}l@{}}conv\_1: 20 filters of shape (8, 20) (nonlinearity: relu)\\
    pool\_1: (2, 2) max pooling\\
    conv\_2: 20 filters of shape (8, 20) (nonlinearity: relu)\\
    pool\_2: (2, 2) max pooling\\
    dense\_1: 200 units (nonlinearity: relu, 10\% dropout)\\
    dense\_2: 10 units (nonlinearity: softmax, 10\% dropout)    
    \end{tabular} \\ \hline
    \textbf{Chirplets (chirp2class)}, shape (80, 110) & \begin{tabular}[x]{l@{}l@{}l@{}l@{}l@{}l@{}}conv\_1: 20 filters of shape (8, 20) (nonlinearity: relu)\\
    pool\_1: (2, 2) max pooling\\
    conv\_2: 20 filters of shape (8, 20) (nonlinearity: relu)\\
    pool\_2: (2, 2) max pooling\\
    dense\_1: 200 units (nonlinearity: relu, 10\% dropout)\\
    dense\_2: 10 units (nonlinearity: softmax, 10\% dropout)    
    \end{tabular} \\ \hline  
  \end{tabular}
  \caption {CNN topologies for the 3 different inputs.}
  \end{table}
\end{center}

\newpage


In all experiments, a given topology is always initialized using the same set of random parameters, unless specified otherwise. The value * (resp. 0) after the name of the net refers to the pretrained net (resp. random initialization).

\section{BIRD Audio2chirp - Chirplet encoder}

The chirp encoder, aka \textit{audio2chirp}, aims at training a net to get a Chirplet-like representation. It is a simple CNN taking audio as input, Chirplets as output and minimizing the square error. It converges easily in 180 epochs. The topology of the audio2chirp net is given Tab. 4.

\begin{center}
\begin{table}[h!]
  \begin{tabular}{| c | c |}
    \hline
    \textbf{Input} & \textbf{Topology} \\ \hline
    \textbf{Audio}, shape (1, 11025) &
    \begin{tabular}[x]{l@{}l@{}l@{}l@{}l@{}l@{}l@{}l@{}}conv\_1: 40 filters of shape (1, 1001) (nonlinearity: relu)\\
    pool\_1: (1, 4) max pooling\\
    conv\_2: 40 filters of shape (1, 501) (nonlinearity: relu)\\
    pool\_2: (1, 4) max pooling\\    
    conv\_3: 40 filters of shape (1, 101) (nonlinearity: relu)\\
    pool\_3: (1, 4) max pooling\\
    dense\_1: 8800 units (nonlinearity: relu, 10\% dropout)\\
    reshape\_1: 8800 -> (80, 110)    
    \end{tabular} \\ \hline    
  \end{tabular}
  \caption {CNN topology of the chirp encoder (\textit{audio2chirp}).}
  \end{table}
\end{center}

\newpage
\section{BIRD: training, dev and testing curves of the different CNNs}


\begin{figure}[h!]
    \centering
    \includegraphics[width=1.1\linewidth]{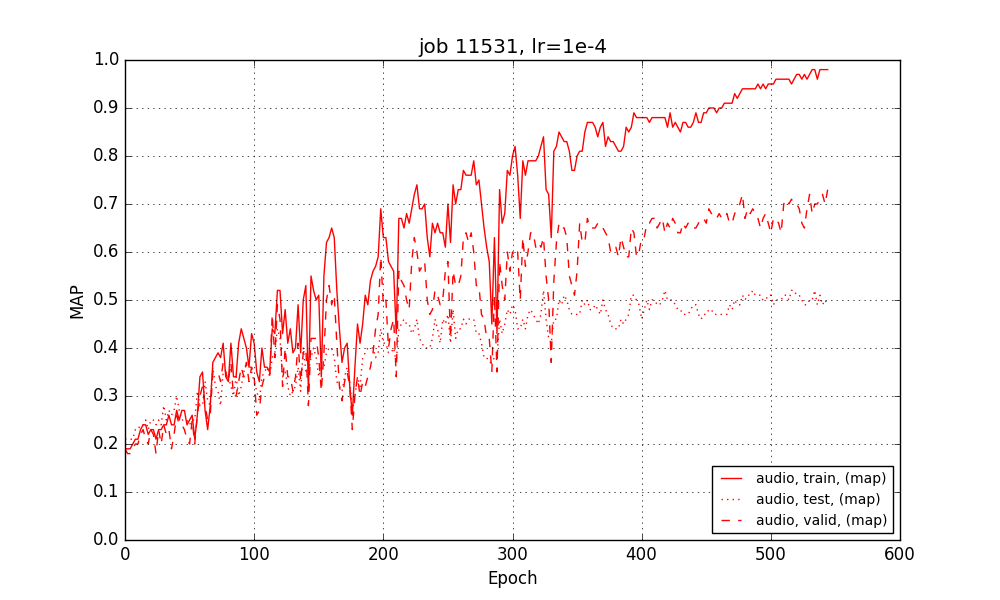}
    \caption{Training stacked random initialized CNNs: audio2chirp(0) and chirp2class(0).}
    \label{fig:my_label31}
\end{figure}

\begin{figure}[h!]
    \centering
    \includegraphics[width=1.1\linewidth]{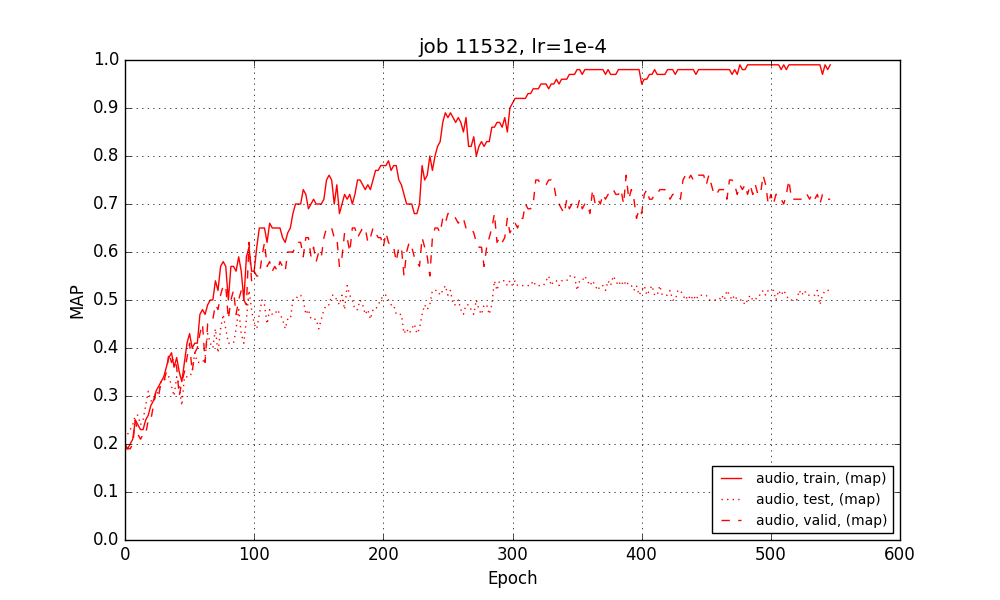}
    \caption{Training stacked pretrained CNNs: audio2chirp(*) and chirp2class(*).}
    \label{fig:my_label32}
\end{figure}

\begin{figure}[h!]
    \centering
    \includegraphics[width=1.1\linewidth]{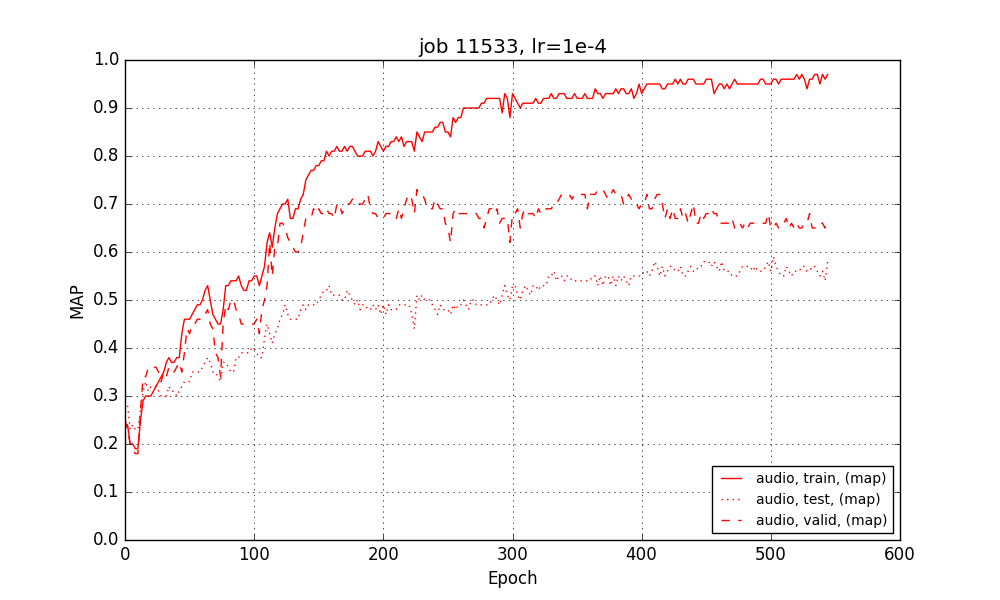}
    \caption{Training stacked pretrained CNNs audio2chirp(*) and chirp2class(*) but freezing chirp2class(*) (no weight update).}
    \label{fig:my_label33}
\end{figure}

\begin{figure}[h!]
    \centering
    \includegraphics[width=1.1\linewidth]{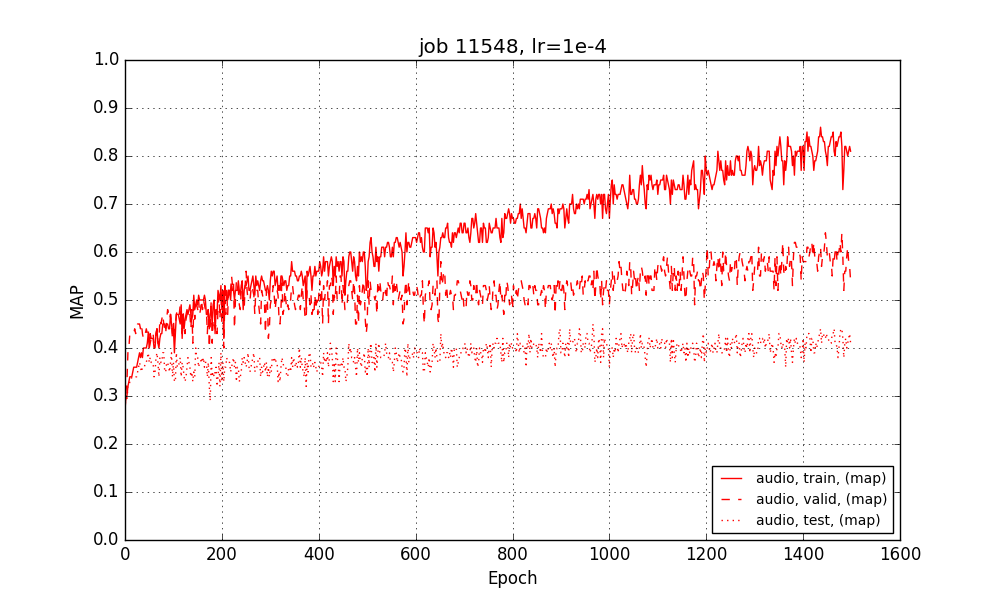}
    \caption{Training stacked pretrained CNNs audio2chirp(*) and chirp2class(*) but freezing audio2chirp(*) (no weight update).}
    \label{fig:my_label48}
\end{figure}

\begin{figure}[h!]
    \centering
    \includegraphics[width=1.1\linewidth]{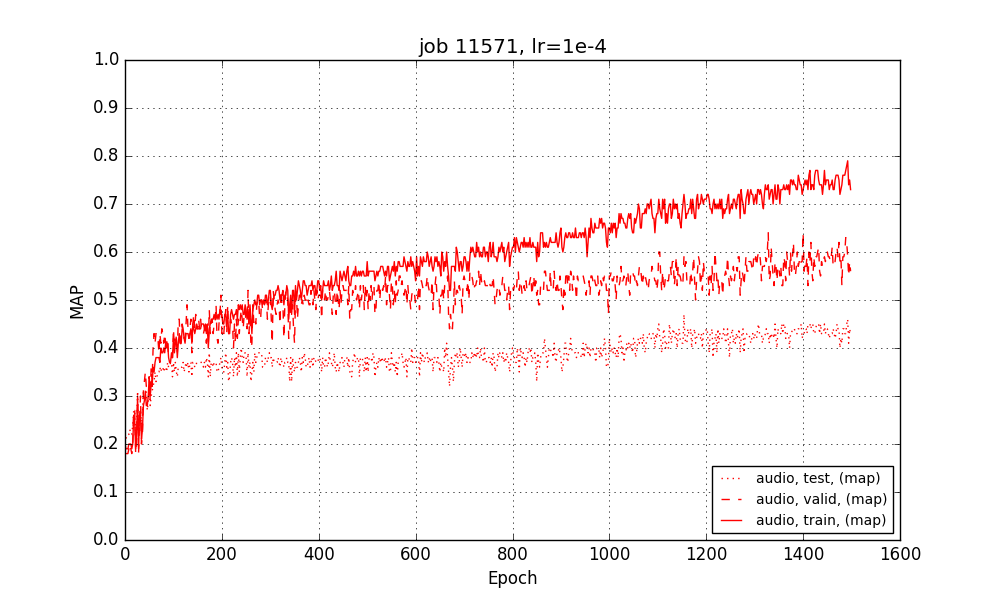}
    \caption{Training stacked CNNs : pretrained audio2chirp(*) and chirp2class(0) and freezing audio2chirp (no weight update).}
    \label{fig:my_label71}
\end{figure}

\begin{figure}[h!]
    \centering
    \includegraphics[width=1.1\linewidth]{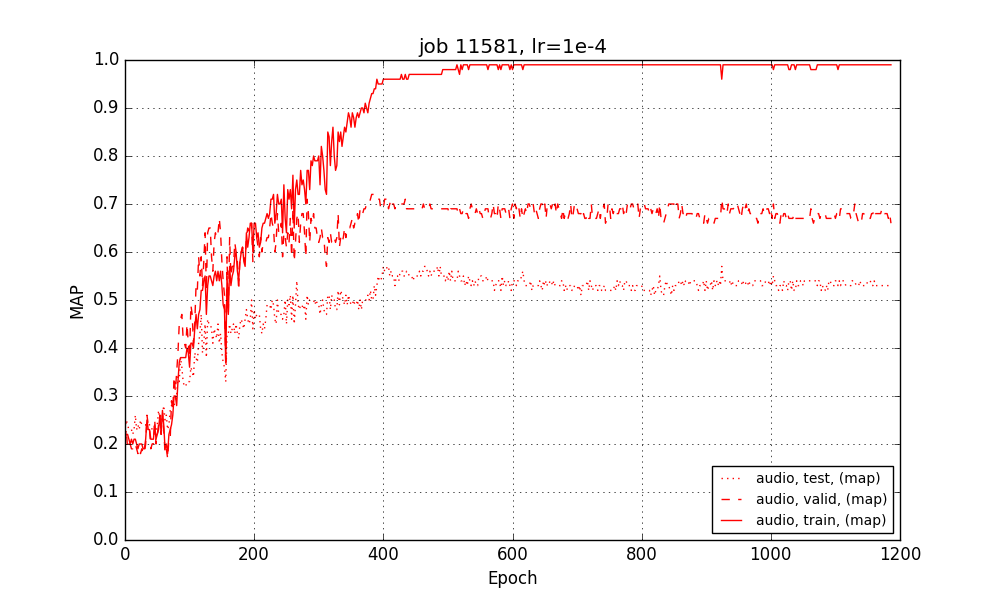}
    \caption{Training stacked CNN from pretrained CNN: Initialized with optimal audio2chirp(*) and chirp2class(0).}
    \label{fig:my_label81}
\end{figure}


\newpage
\section{Experiment on Speech Vowel}

The Tab. \ref{tab:topoCNNTIMIT} gives the topology of the audio2chirp and chirp2class, and stacked models, for these vowel experiments.

In all experiments, each CNN is initialized using the same random seed. The symbol “*” refers to the optimal trained parameters of a net.

\begin{center}
\begin{table}[h!]
  \begin{tabular}{| c | c |}
    \hline
    \textbf{Input} & \textbf{Topology} \\ \hline
    \textbf{audio2chirp}, shape (1, 4960) &
    \begin{tabular}[x]{l@{}l@{}l@{}l@{}l@{}l@{}l@{}l@{}}conv\_1: 40 filters of shape (1, 1001) (nonlinearity: relu)\\
    pool\_1: (1, 4) max pooling\\
    conv\_2: 40 filters of shape (1, 501) (nonlinearity: relu)\\
    pool\_2: (1, 4) max pooling\\
    conv\_3: 40 filters of shape (1, 101) (nonlinearity: relu)\\
    pool\_3: (1, 4) max pooling\\
    dense\_1: 3136 units (nonlinearity: relu, 10\% dropout)\\
    reshape\_1: 3136 -> (64, 49)   
    \end{tabular} \\ \hline
    \textbf{chirp2class}, shape (64, 49) &
    \begin{tabular}[x]{l@{}l@{}l@{}l@{}l@{}l@{}}conv\_1: 20 filters of shape (8, 10) (nonlinearity: relu)\\
    pool\_1: (2, 2) max pooling\\
    conv\_2: 20 filters of shape (8, 10) (nonlinearity: relu)\\
    pool\_2: (2, 2) max pooling\\
    dense\_1: 200 units (nonlinearity: relu, 10\% dropout)\\
    dense\_2: 8 units (nonlinearity: softmax, 10\% dropout)    
    \end{tabular} \\ \hline    
  \end{tabular}
  \caption {CNN topologies for TIMIT vowel experiments.}
  \label{tab:topoCNNTIMIT}
  \end{table}
\end{center}

\begin{figure}[h!]
    \centering
    \includegraphics[width=1.1\linewidth]{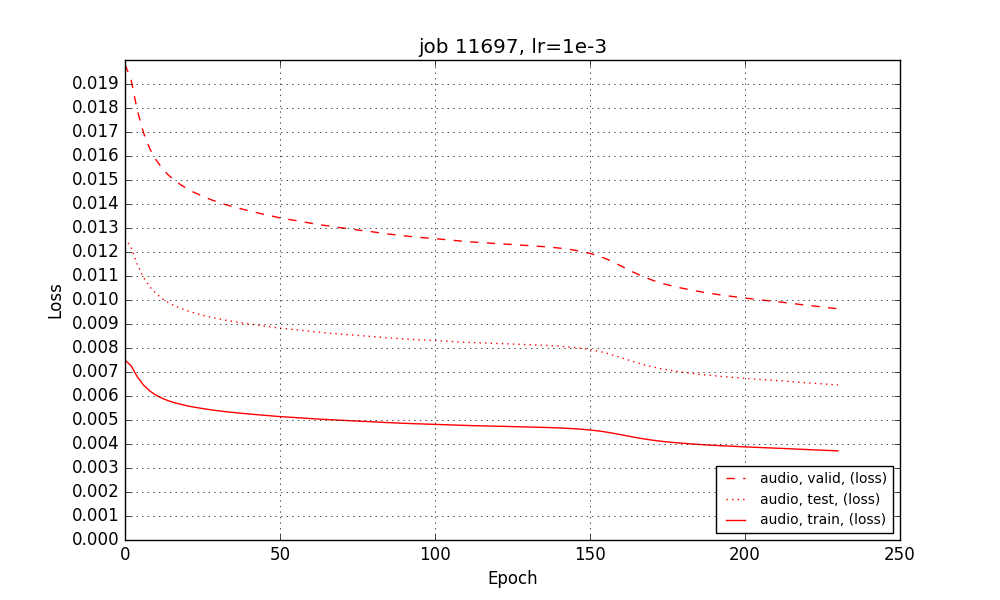}
    \caption{Trained and loss audio2chirp (TIMIT).}
    \label{fig:my_label2}
\end{figure}


\newpage
\section{Algorithm for the Fast Chirplet Transform (FCT)}

\begin{verbatim}
    Algo 1: Chirplet Generation
    INPUT: F0,F1,Fs,sigma,p
    OUTPUT: coefficients_upward,coefficients_downward
    if(p):
        w=cos(2*pi*((F1-F0)/((p+1)*sigma**p)*t**p+F0)*t)
    else:
        w=cos(2*pi*((F0*(F1/F0)**(t/sigma)-F0)*sigma/log(F1/F0)))
    coefficients_upward=w*exp(-((t-\sigma/2.0)**2)/(2*sigma**2))
    coefficients_downward=flipud(coefficients_upward).
\end{verbatim}

\begin{verbatim}
    Algo 2: Chirplet Filter-Bank Generation
    INPUT: J, Q, Fs, sigma, p
    lambdas            = 2.0**(1+arrange(J*Q)/float(Q))
    start_frequencies  = (Fs /lambdas)/2.0
    end_frequencies    = Fs /lambdas
    distances          = 2.0*d/flipud(lambdas)
    filters=list()
    for f0,f1,d in zip(start_frequencies,end_frequencies,distances):
        filters.append(chirplet(Fs,f0,f1,d,p))
    return filters.
\end{verbatim}

\newpage
\section{The Python code for the Fast Chirplet Transform (FCT)}

This code, in GPL licence (c) DYNI team, is in Github :\\ \url{https://github.com/DYNI-TOULON/fastchirplet.git }.

\begin{lstlisting}[language=Python,breaklines=true ]
import librosa
import os
import numpy as np
from pylab import *
import sys
from numpy.lib import pad

class Chirplet:

	"""smallest time bin among the chirplet"""
	global smallest_time_bins

	def __init__(self,samplerate,F0,F1,sigma,polynome_degree):

		"""lowest frequency where the chirplet is applied"""
		self.min_frequency  = F0

		"""highest frequency where the chirplet is applied"""
		self.max_frequency = F1

		"""samplerate of the signal"""
		self.samplerate = samplerate

		"""duration of the chirplet"""
		self.time_bin = sigma/10

		"""degree of the polynome to generate the coefficients of the chirplet"""
		self.polynome_degree = polynome_degree

		"""coefficients applied to the signal"""
		self.filter_coefficients = self.calcul_coefficients()


	def calcul_coefficients(self):
		"""calculate coefficients for the chirplets"""
		# print(self._samplerate)
		t = linspace(0,self.time_bin,int(self.samplerate*self.time_bin))
		if(self.polynome_degree):
			w = cos(2*pi*((self.max_frequency-self.min_frequency)
				/((self.polynome_degree+1)*self.time_bin**self.polynome_degree)
				*t**self.polynome_degree+self.min_frequency)*t)
		else:
			w = cos(2*pi*((self.min_frequency*(self.max_frequency/self.min_frequency)
				**(t/self.time_bin)-self.min_frequency)*self.time_bin
			/log(self.max_frequency/self.min_frequency)))
		
		coeffs = w*hanning(len(t))**2

		return coeffs

	def smooth_up(self,input_signal,sigma,end_smoothing):
	#generate fast fourier transform from a signal and smooth it

		new_up = build_fft(input_signal,self.filter_coefficients,sigma)
		return fft_smoothing(fabs(new_up),end_smoothing)


def compute(input_signal,save=False,duration_last_chirplet=1.01,
	num_octaves=5,num_chirps_by_octave=10,polynome_degree=3,end_smoothing=0.001):
	"""main function. Fast Chirplet Transform from a signal"""

	data, samplerate = librosa.load(input_signal,sr=None)

	size_data = len(data)

	nearest_power_2 = 2**(size_data-1).bit_length()

	data = np.lib.pad(data,(0,nearest_power_2-size_data),
		'constant',constant_values=0)

	chirplets = init_chirplet_filter_bank(samplerate,duration_last_chirplet,num_octaves,num_chirps_by_octave,polynome_degree)

	chirps = apply_filterbank(data,chirplets,end_smoothing)

	chirps = resize_chirps(size_data,nearest_power_2,chirps)

	if save:
		if not os.path.exists("csv"):
			os.makedirs("csv")
		np.savetxt("csv/"+os.path.basename(input_signal).split('.')[0]+'.csv',chirps, delimiter=",")

	return chirps

def resize_chirps(size_data,nearest_power_2,chirps):
	size_chirps = len(chirps)
	ratio = size_data/nearest_power_2
	size = int(ratio*len(chirps[0]))

	tabfinal = np.zeros((size_chirps,size))
	for i in range(0,size_chirps):
		tabfinal[i]=chirps[i][0:size]
	return tabfinal

def init_chirplet_filter_bank(samplerate,duration_last_chirplet,num_octaves,num_chirps_by_octave,p):
	"""generate all the chirplets from a given sample rate"""

	lambdas            = 2.0**(1+arange(num_octaves*num_chirps_by_octave)/float(num_chirps_by_octave))
	#Low frequencies for a signal
	start_frequencies  = (samplerate /lambdas)/2.0
	#high frequencies for a signal
	end_frequencies    = samplerate /lambdas
	durations          = 2.0*duration_last_chirplet/flipud(lambdas)
	Chirplet.smallest_time_bins = durations[0]
	chirplets=list()
	for f0,f1,duration in zip(start_frequencies,end_frequencies,durations):
		chirplets.append(Chirplet(samplerate,f0,f1,duration,p))
	return chirplets

def apply_filterbank(input_signal,chirplets,end_smoothing):
	"""generate list of signal with chirplets"""
	result=list()
	for chirplet in chirplets:
		result.append(chirplet.smooth_up(input_signal,6,end_smoothing))
	return array(result)



def fft_smoothing(input_signal,sigma):
	"""smooth the fast transform fourier"""
	size_signal = input_signal.size
	#shorten the signal
	new_size = int(floor(10.0*size_signal*sigma))
	half_new_size = new_size//2

	fftx = fft(input_signal)
	short_fftx = []
	for ele in fftx[:half_new_size]:
		short_fftx.append(ele)
	for ele in fftx[-half_new_size:]:
		short_fftx.append(ele)

	apodization_coefficients = generate_apodization_coefficients(half_new_size,sigma,size_signal)
	#apply the apodization coefficients
	short_fftx[:half_new_size] *= apodization_coefficients
	#apply the apodization coefficients in a reverse list
	short_fftx[half_new_size:] *= flipud(apodization_coefficients)
	realifftxw = real(ifft(short_fftx))
	return realifftxw
	
def generate_apodization_coefficients(num_coeffs,sigma,size):
	"""generate apodization coefficients"""
	apodization_coefficients = arange(num_coeffs)
	apodization_coefficients = apodization_coefficients**2
	apodization_coefficients = apodization_coefficients/(2*(sigma*size)**2)
	apodization_coefficients = exp(-apodization_coefficients)
	return apodization_coefficients

def fft_based(input_signal,h,boundary=0):
	M=h.size
	half_size = M//2
	if(boundary==0):#ZERO PADDING
		input_signal=pad(input_signal,(half_size,half_size),'constant',constant_values=0)
		h=pad(h,(0,input_signal.size-M),'constant',constant_values=0)
		newx=ifft(fft(input_signal)*fft(h))
		return newx[M-1:-1]
	elif(boundary==1):#symmetric
		input_signal=concatenate([flipud(input_signal[:half_size]),input_signal,flipud(input_signal[half_size:])])
		h=pad(h,(0,x.size-M),'constant',constant_values=0)
		newx=ifft(fft(input_signal)*fft(h))
		return newx[M-1:-1]
	else:#peridic
		return real(roll(ifft(fft(input_signal)*fft(h,input_signal.size)),-half_size))


def build_fft(input_signal,filter_coefficients,n=2,boundary=0):
	"""generate fast transform fourier by windows"""
	M=filter_coefficients.size
	#print(n,boundary,M)
	half_size = M//2
	signal_size = input_signal.size
	#power of 2 to apply fast fourier transform
	windows_size=int(2**ceil(log2(M*(n+1))))
	number_of_windows=floor(signal_size//windows_size)
	if(number_of_windows==0):
		return fft_based(input_signal,filter_coefficients,boundary)

	output=empty_like(input_signal)
	#pad with 0 to have a size in a power of 2
	zeropadding = pad(filter_coefficients,(0,windows_size-M),'constant',constant_values=0)

	h_fft=fft(zeropadding)

	#to browse the whole signal
	current_pos=0

	#apply fft to a part of the signal. This part has a size which is a power of 2
	if(boundary==0):#ZERO PADDING
		#window is half padded with since it's focused on the first half
		window = input_signal[current_pos:current_pos+windows_size-half_size]
		zeropaddedwindow = pad(window,(len(h_fft)-len(window),0),'constant',constant_values=0)
		x_fft=fft(zeropaddedwindow)
	elif(boundary==1):#SYMMETRIC
		window = concatenate([flipud(input_signal[:half_size]),input_signal[current_pos:current_pos+windows_size-half_size]])
		x_fft=fft(window)
	else:
		x_fft=fft(input_signal[:windows_size])

	output[:windows_size-M]=ifft(x_fft*h_fft)[M-1:-1]
	current_pos+=windows_size-M-half_size

	#apply fast fourier transofm to each windows 
	while(current_pos+windows_size-half_size<=signal_size):
		x_fft=fft(input_signal[current_pos-half_size:current_pos+windows_size-half_size])
		output[current_pos:current_pos+windows_size-M]=real(ifft(x_fft*h_fft)[M-1:-1])

		current_pos+=windows_size-M

	#apply fast fourier transform to the rest of the signal
	if(windows_size-(signal_size-current_pos+half_size)<half_size):
		window = input_signal[current_pos-half_size:]
		zeropaddedwindow = pad(window,(0,int(windows_size-(signal_size-current_pos+half_size))),'constant',constant_values=0)
		x_fft=fft(zeropaddedwindow)
		output[current_pos:]=real(roll(ifft(x_fft*h_fft),half_size)[half_size:half_size+output.size-current_pos])
		output[-half_size:]=convolve(input_signal[-M:],filter_coefficients,'same')[-half_size:]
	else:
		window = input_signal[current_pos-half_size:]
		zeropaddedwindow = pad(window,(0,int(windows_size-(signal_size-current_pos+half_size))),'constant',constant_values=0)
		x_fft=fft(zeropaddedwindow)
		output[current_pos:]=real(ifft(x_fft*h_fft)[M-1:M+output.size-current_pos-1])
	return output
\end{lstlisting}

\end{document}